\documentclass[useAMS,usenatbib,aps,floatdix,nofootinbib]{mn2e}

\usepackage{graphics}
\usepackage{amssymb,amsmath}
\usepackage{amsmath}
\usepackage{psfrag}
\usepackage{graphicx}

\newcommand{\f}{\frac}

\newcommand{\be}{\begin{equation}}      
\newcommand{\ee}{\end{equation}}      
      
\newcommand{\bef}{\begin{figure}}      
\newcommand{\eef}{\end{figure}}      
\newcommand{\bea}{\begin{eqnarray}}    
\newcommand{\eea}{\end{eqnarray}}      
  
\def\spose#1{\hbox to 0pt{#1\hss}}
\def\ltapprox{\mathrel{\spose{\lower 3pt\hbox{$\mathchar"218$}}
\raise 2.0pt\hbox{$\mathchar"13C$}}}
\def\gtapprox{\mathrel{\spose{\lower 3pt\hbox{$\mathchar"218$}}
\raise 2.0pt\hbox{$\mathchar"13E$}}}
\def\inapprox{\mathrel{\spose{\lower 3pt\hbox{$\mathchar"218$}}
\raise 2.0pt\hbox{$\mathchar"232$}}}

\def\bse{\begin{subequations}}
\def\ese{\end{subequations}}

\def\lsim{\raise 0.4ex\hbox{$<$}\kern -0.8em\lower 0.62ex\hbox{$\sim$}} 
\def\gsim{\raise 0.4ex\hbox{$>$}\kern -0.7em\lower 0.62ex\hbox{$\sim$}}

\def\f0N{f_0^{(N)}}
\def\bec{\begin{center}}
\def\eec{\end{center}}

\title[Exponents of non-linear clustering in one dimension]
{Exponents of non-linear clustering in scale-free one dimensional 
cosmological simulations}
\author[D. Benhaiem, M. Joyce and F. Sicard]
{David Benhaiem${^1}$, Michael Joyce${^1}$ and Fran\c{c}ois Sicard${^{2}}$ \\
$^1$Laboratoire de Physique Nucl\'eaire et Hautes \'Energies, 
Universit\'e Pierre et Marie Curie - Paris 6, CNRS IN2P3 UMR 7585, \\  4 Place Jussieu, 75752 Paris Cedex 05, France\\
$^{2}$Laboratoire Interdisciplinaire Carnot de Bourgogne,
UMR 6303 CNRS-Universit\'e de Bourgogne, \\
9 Avenue A. Savary, BP 47 870, F-21078 Dijon Cedex, France
}

\begin{document}

\date{\today}

\maketitle

\begin{abstract}
One dimensional versions of dissipationless cosmological N-body simulations have been shown 
to share many qualitative behaviours of the three dimensional problem. Their interest 
lies in the fact that they can resolve a much greater range of time and length scales,
and admit exact numerical integration. We use such models
here to study how non-linear clustering depends on initial conditions and cosmology. More specifically, we 
consider a family of models which, like 
the three dimensional EdS model, lead for power-law initial conditions
to self-similar clustering characterized in the strongly non-linear regime 
by power-law behaviour of the two point correlation function. We study how 
the corresponding exponent $\gamma$ depends on the initial conditions, 
characterized by the exponent $n$ of the power spectrum of initial fluctuations,
and on a single parameter $\kappa$ controlling the rate of expansion.  
The space of initial conditions/cosmology divides very clearly into 
two parts:  (1) a region in which $\gamma$ depends strongly on 
both $n$ and $\kappa$ and where it agrees very well with a 
simple generalisation of the so-called stable clustering hypothesis
in three dimensions, and (2) a region in which $\gamma$
is more or less independent of both the  spectrum and the expansion 
of the universe. The boundary in $(n, \kappa)$ space dividing
the ``stable clustering" region from the ``universal" region
is very well approximated by a ``critical" value of the predicted
stable clustering exponent itself.  We explain how this division
of the $(n, \kappa)$  space can be understood as a simple physical criterion 
which might indeed be expected to control the validity of the stable 
clustering hypothesis.
We compare and contrast our findings to results in three dimensions, and discuss 
in particular the light they may  throw on the question of ``universality" of 
non-linear clustering in this context.
\end{abstract}


\section{introduction}


Cosmological $N$ body simulations are the primary instrument used to make theoretical predictions 
for structure formation in current models of the universe. Analytical understanding of the non-linear 
regime, crucial for many non-trivial tests of these models, remains very poor despite the many
phenomenological results derived in simulations. Better understanding of the physics of non-linear
clustering could help to better constrain and control cosmological simulations, which despite their 
ever increasing power and sophistication are still subject to questions about their accuracy and
resolution. In this work we use simplified one dimensional  (1D) ``toy models" of the full three dimensional 
(3D) problem  to study non-linear structure formation. More specifically we study, using a family of
such models,  how the properties of non-linear clustering depend on both initial conditions and 
cosmology. Our main result is that, for cold initial conditions,  we observe that the large phase
space of initial conditions and cosmology we explore breaks up into two parts: on the
one hand, a region in which there is an apparent   ``universality'' of the non-linear clustering, i.e., in 
which it depends very weakly both on initial conditions and on the cosmological expansion; and
on the other hand, a region where the non-linear clustering depends strongly on both the initial
conditions and expansion history.  In the latter region, the exponent characterizing the two
point correlations is in excellent agreement with a simple analytical prediction which follows
from the so-called stable clustering hypothesis, appropriately generalized to this class
of models.  Further the apparently well localized boundary  dividing the phase space 
into these two disjoint regions can in fact be characterized using the stable clustering 
prediction, in terms of a critical value of the corresponding exponent.  The non-linear 
clustering appears to be strictly universal in the limit in which there is no expansion,
but also very well approximated for steeper initial spectra when there is 
expansion.  We compare our results with those in three dimensions, and 
discuss how our results may provide a framework for understanding the much 
discussed apparent ``universality" of halo profiles in this context. 
 
The exploration of 1D models as a tool for understanding the physics
of non-linear structure formation in a cosmological context goes back at least as
far as the work of  \cite{melott_prl_1982,melott_1d_1983} for the case of 
hot dark matter models. Different initial conditions and variants of the model 
have been discussed by a number of other authors  
\citep{yano+gouda, miller+rouet_2002, aurell+fanelli_2002a, miller+rouet_2006, valageasOSC_2,
miller_etal_2007, agmjfs_pre2009, miller+rouet_2010a}.  
There is also an extensive literature on the statistical mechanics of 
finite one dimensional self-gravitating systems (see e.g. \cite{joyce+worrakitpoonpon_2012} and 
references therein), as well as some studies which use this case to explore issues
in  cosmological N body simulations:  \cite{binney_discreteness} uses it to probe
discreteness effects, and \cite{schulz_etal_2012} the issue of universality in
cold collapse.

The study reported here is a direct continuation of that in \cite{joyce+sicard_2011}.
This latter paper showed, for the three 1D models studied in the previous literature,
the very strong qualitative similarities in the temporal development of clustering to that 
in the 3D case. For cold cosmological like initial conditions clustering is hierachical
and driven by the linear amplification of fluctuations, leading, for the case of 
power law initial conditions, to self similar evolution of the non-linear
correlations.  The exponent characterising the non-linear clustering 
was, further, found to be in very good agreement with that predicted
by the so-called ``stable clustering" hypothesis. In this paper we extend
this study to a one parameter family of models which allows
continuous interpolation between the static model and the
two expanding models previously considered. This allows us
to probe fully the extent of validity of stable clustering, where
it breaks down and what happens in this case. The study 
reveals that there is a fairly abrupt switchover from the 
validity of stable clustering to an apparent universality in the 
clustering. 

In the next section we define the class of 1D models we study, 
discussing also their relation to 3D cosmological models. This
presentation improves and simplifies that given 
in \cite{joyce+sicard_2011}.
For the family of scale-free models we focus on, we derive, more
rigorously than in \cite{joyce+sicard_2011}, the
prediction for the behaviour of an isolated structure and use
it to derive the exponents predicted for the non-linear 
two point correlation when the stable clustering hypothesis
is made.  In the following section we discuss our numerical
simulations, and in particular how we calibrate them using
``exact" simulations. The next section gives our results,
and in the last section we discuss their interpretation
and possible relation to the three dimensional problem.

\section {A family of 1D scale-free models}  

The class of 1D models we study are defined as follows.
We take the equations of motion {\it in comoving coordinates}  
of 3D cosmological $N$ body simulations,  and simply
replace the 3D Newtonian gravitational interaction
with the 1D gravitational interaction.  This gives 
1D equations of motion
\begin{equation}
\frac{d^2 x_i}{dt^2} +
2H \frac{d x_i}{dt}= -\frac{g}{a^3}
\sum^J_{j\neq i}
\textrm{sgn}(x_i - x_j) 
\label{1Deom1}
\end{equation}
where $x_i$ are the particle positions on the line. The 
superscript `J' in the force term indicates that the sum, which 
extends over an infinite distribution  of masses with non-zero 
mean mass density,  is  regularized,  just  as in 3D, by 
subtraction of the contribution of the mean mass density, i.e., 
the force term is sourced only by fluctuations about the 
mean density. As we will discuss further below this
sum can be written explicitly in a simple manner.

The coupling $g$ is the 1D analogy of Newton's constant 
(multiplied by the particle mass). To establish a more direct relation 
between the 1D and 3D models,  one can consider the ``particles"
in the 1D system to represent infinite parallel sheets of zero
thickness embedded in 3D. In this case one has the identification 
$g=2 \pi G \Sigma$ where $\Sigma$ is the surface 
mass density of the sheets. The mean (comoving) mass 
density $\rho_0$ of the corresponding 3D universe and the 
number density $n_0$ of the 1D system are then related 
as $\rho_0=\Sigma n_0$, and therefore 
\begin{equation}
gn_0= 2 \pi G \rho_0\,.
\label{g-Grelation}
\end{equation}
For any given 3D cosmological model which provides
a scale factor $a(t)$ and associated Hubble constant $H(t)$
we then have a {\it unique} corresponding 1D model.
The case $a(t)=1$ (and $H(t)=0$) on the other hand, defines 
a model analogous to a 3D universe without expansion
(which is well defined but does not correspond to a 
cosmological model derived from general relativity). 

The one parameter family of models we will study is 
a {\it continuous interpolation between the 1D model 
obtained using the 3D EdS cosmology, and  the 
static model}. To  define them precisely, and motivate 
their choice, it is convenient to change time variable 
in  Eq.~(\ref{1Deom1}) defining $\tau=\int \frac{dt}{a^{3/2}}$. 
This gives the equations of motion in the
form 
\begin{equation}
\frac{d^2 x_i}{d\tau^2} +
\Gamma (\tau) \frac{d x_i}{d\tau}= -{g}
\sum^J_{j\neq i}
\textrm{sgn}(x_i - x_j) \,,
\label{1Deom2}
\end{equation}
i.e., in which all effects of the cosmology appear only as a fluid
type damping term, with
\begin{equation}
\Gamma =\frac{1}{2} a^{3/2} H =\frac{1}{2a} \frac{da}{d\tau}\,.
\label{Gamma-a-relation}
\end{equation}
For a 3D EdS universe with comoving
mass density $\rho_0$, i.e., 
$a=(t/t_0)^{2/3}$ where $t_0=1/\sqrt{6\pi G \rho_0}$,
we obtain a $\Gamma$ which is {\it independent of time}
and given by 
\begin{equation}
\Gamma = \frac{1}{3t_0}=\sqrt{2\pi G \rho_0/3}=  \sqrt{g n_0/3} 
\label{Gamma-EdS}
\end{equation}
where we have used Eq.~(\ref{g-Grelation}) to derive the
last equality. In a $\Lambda$CDM cosmology,  on
the other hand,  $\Gamma$ is approximately constant 
through the matter dominated era and increases 
slightly at late times.  

We consider here simply the set of 1D models
in which $\Gamma$ takes any constant value,
and the focus of our study is the dependence of
non-linear clustering on this parameter, as
well as on the initial conditions.
$\Gamma$ can simply be considered as 
a control parameter for understanding 
the role of expansion in the determination
of the properties of the non-linear clustering.
While most previous studies have considered either
the EdS model (also known as the ``quintic" 
model for reasons which will be recalled below), or
the static model, one other model in this family, corresponding to 
the case  $\Gamma=\sqrt{gn_0}$, introduced 
in \cite{miller+rouet_2002}, has been studied 
in \cite{miller+rouet_2006, miller_etal_2007, miller+rouet_2010a}
as well as in \cite{joyce+sicard_2011}. 

We note, using Eq.~(\ref{Gamma-a-relation}), that
this class of models is obtained by taking the 
functional dependence of the 3D EdS expansion law,
but allowing a freedom in the normalisation of the 
expansion rate to the matter density, i.e., we 
can obtain these models taking 
$H^2= \kappa^2 {8\pi G \rho_0/3a^3}$ where
$\kappa$ is a positive constant.
Eq.~(\ref{Gamma-a-relation}) then still
holds, and thus $a=e^{2 \Gamma \tau}$,
but instead of (\ref{Gamma-EdS}) we
have, using Eq.~(\ref{g-Grelation}), that
\begin{equation} 
\Gamma=\kappa \sqrt{g n_0/3}\,.
\end{equation} 
In terms of 3D cosmological models 
the case $\kappa >1$, i.e., $\Gamma > \Gamma_{\rm EdS}$, 
can be considered to be 
a model in which there is not just ``ordinary" 
matter but an additional pressureless component 
of the energy density which does not cluster, e.g., a 
homogeneous scalar field with appropriate
equation of state (see e.g. \cite{ferreira+joyce_1998}
and references therein).  Equivalently one
can consider that  the $\kappa \neq 1$
models are obtained using the ``correct"  
normalisation for the Hubble constant, 
$H^2={8\pi G \rho_0/3}$, but choosing the 
1D coupling freely instead
of  imposing the relation Eq.~(\ref{g-Grelation}).
It is important to note that, given that 3D 
Hubble law is imposed  ``by hand" on the
1D model, there is in principle no correct
value for $\Gamma$. Our point of view here is 
simply to use $\Gamma$ (or $\kappa$) as a control parameter 
in trying to understand the role of expansion in non-linear
structure formation. In other words we are 
seeking to improve our {\it qualitative} understanding 
only through the study of these toy models.

\section {Stable clustering predictions}  

In the context of the problem of structure formation
the family of models we study has the interest of being, 
like EdS models and static models in 3D, scale-free: 
the expansion introduces no additional time or length 
scales.  This has as a consequence that, for cold
particles with initial density fluctuations with a 
power law spectrum, one expects to obtain,
at sufficiently long times, temporal 
evolution which
is ``self-similar", i.e., the evolution in time
of the spatial correlation functions is equivalent
to a simple time dependent rescaling of 
the spatial variables. 
The reason why this occurs is that there 
is, in this case,  only a single physically
relevant length scale in the problem: the scale 
at which  fluctuations are of order one, of which the evolution
is predicted by linear theory\footnote{In the discrete
problem there is at least one additional length
scale --- the initial interparticle spacing --- but physical
(continuum limit) results should not depend on it.}. 
Further if the clustering in the strongly non-linear regime does not
depend on this scale one would expect to obtain 
a scale-free behaviour of spatial correlations 
in this regime. Such power law behaviour has 
been observed in both expanding 
(see, e.g.,  \cite{efstathiou_88, smith} and 
references therein) and static 3D simulations
\citep{sl1, sl2, sl3} for a range of power law initial 
conditions. 3D simulations, however, can show
such behaviour over a very limited spatial
range, making it difficult to establish if it is
associated with a true scale invariance
of the non-linear clustering. The study
of 1D models in the family we are considering
\citep{miller+rouet_2006, miller_etal_2007, miller+rouet_2010a, joyce+sicard_2011},
with much greater spatial resolution and numerical
accuracy,  shows very convincing evidence that
these models do indeed give rise to power-law clustering 
indicative of truly scale-invariant clustering. 

A central problem of non-linear structure formation
is then that of understanding how the exponent
(or possibly exponents) characterizing the non-linear clustering
is determined. In 3D EdS cosmology Peebles (see \cite{peebles})
proposed many years ago a simple explanation of how such
scale invariant clustering might arise:
if highly
non-linear structures (of all different sizes) are 
supposed to evolve essentially independently 
after their formation, they will virialize and 
remain stable in physical coordinates. From
this one may derive an exponent which
depends only on the exponent in the power
spectrum of initial density fluctuations. 
We derive here now, more rigorously than 
the heuristic derivation given in \cite{joyce+sicard_2011},
the prediction of such a stable clustering hypothesis
for the 1D models we study. To do so we
simply derive and then analyse the equations 
of motion of a finite sub-system.  

\subsection{Equations of motion of a finite subsystem}

In the simulations which we consider here, we 
start, as in 3D cosmological simulations, from a
particle configuration generated by applying
small displacements to an {\it infinite} perfect 
lattice configuration. At any moment in time such a 
configuration can be fully specified by giving 
the displacements $u_i$ of particles from the 
sites of a perfect lattice. The position of
the $i$-th particle with respect to some 
arbitrary origin, is $x_i=i \lambda + u_i$, 
where $\lambda=n_0^{-1}$. In a particle labelling 
with  
$x_i \geq x_{i-1}$ for all $i$, the displacements 
are uniquely defined,  and such that no two particles
cross when they are applied to the lattice.
In this case it has been shown \citep{agmjfs_pre2009}
that the force on the particle $i$, given 
by the sum on the right side of  Eq.~(\ref{1Deom2}),
with an appropriate specification of the regularisation,   
has the simple {\it exact} expression 
\begin{equation}
-g\sum^J_{j\neq i} \textrm{sgn}(x_i - x_j) = 2gn_0 (u_i -\langle u \rangle)
\label{1D-exact-force-disp}
\end{equation}
where
$\langle u \rangle$ is the average particle displacement. 
This result has been shown for an infinite
system with statistically translationally invariant 
displacements in \cite{agmjfs_pre2009}, with the sole
assumption of decay of correlations of the displacements 
at asymptotically large scales. The same result has
been shown to hold in \cite{miller+rouet_2010b} 
for the case of an infinite {\it periodic} system, using 
an appropriate regularisation of an Ewald summation. 
In this case, as we will now show explicitly, the 
expression (\ref{1D-exact-force-disp}) can be
written as 
\begin{equation}
-g\sum^J_{j\neq i} \textrm{sgn}(x_i - x_j) = -g [N_L(i)-N_R(i)] + 2gn_0[x_i - \langle x \rangle_p] 
\label{1D-force-periodic}
\end{equation}
where $N_L(i)$ ($N_R(i)$) is the number of particles
in the periodic cell to the left (right) of particle $i$, 
and $\langle x \rangle_p$ is the average particle
position, i.e., centre of mass of the particles in
the cell\footnote{Contrary to what is stated 
in \cite{miller+rouet_2010b}, the expressions 
for the force in the two papers are in agreement.}.

Let us consider now any {\it finite} subsystem ${\cal S}$
of the infinite system defined simply by the $N$
particles, with $i=1...N$ say, in some interval 
$[x_0, x_0+L]$ at a given instant in time,
Substituting $u_i= x_i - i \lambda$ in (\ref{1D-exact-force-disp}),
and noting that the number of particles in the interval
to the left (right) of the $i$th particle is given
by $N_L(i)=i-1$ ($N_R(i)=N-i$), we obtain
that the force on the particle $i$ may be 
written 
\begin{equation}
F(x_i) = -g [N_L(i)-N_R(i)] + 2gn_0[x_i - \langle x \rangle_{\cal S}] - 2gn_0 [\langle u \rangle
- \langle u \rangle_{\cal S}]
\label{1D-force-subsystem}
\end{equation}
where  $\langle x \rangle_{\cal S}$ is simply the position of the centre of mass 
of the subsystem, and $\langle u \rangle_{\cal S}$ the average displacement 
of the particles in the subsystem. If the subsystem chosen is the
periodic cell of an infinite periodic system, we have $\langle u \rangle
= \langle u \rangle_{\cal S}$ and thus obtain (\ref{1D-force-periodic}).
For any generic subsystem ${\cal S}$, on the other hand, the last term 
on the right in (\ref{1D-force-subsystem}) is the same for all particles 
in the subsystem. Defining $y_i=x_i-\langle x \rangle_{\cal S}$
we thus obtain  the equation of motion for any particle in the 
subsystem, relative to the CM of the subsystem, as 
\begin{equation}
\frac{d^2 y_i}{d\tau^2} +
\Gamma \frac{d y_i}{d\tau}= -{g}
\sum_{j\neq i,  j \in {\cal S}}
\textrm{sgn}(y_i - y_j) + 2gn_0 y_i
\label{1Deom-finite-finale}
\end{equation}
where the sum extends over the particles in the subsystem.
This remains valid for as long as no other particle outside the system
crosses the particles at its extremities. Eq.~(\ref{1Deom-finite-finale}) 
are thus exactly the equations of motion of a subsystem as 
long as it is isolated in this sense.

\subsection{Definition of ``quasi-physical" coordinates}

For $\Gamma \neq 0$ we can transform to the coordinates
\begin{equation}
\tilde{t}=\frac{3}{\Gamma} e^{\Gamma \tau / 3}, 
\quad r_i= e^{2 \Gamma \tau /3} y_i
\label{physical coordinates}
\end{equation}
in which Eq.~(\ref{1Deom-finite-finale}) becomes
\begin{equation}
\frac{d^2 r_i}{d \tilde{t}^2} = -{g}
\sum_{j\neq i,  j \in {\cal S}}
\textrm{sgn}(r_i - r_j) + \frac{2}{\tilde{t}^2}[1+ \frac{9gn_0}{\Gamma^2}]r_i\,.
\label{1Deom-finite-phys-coords}
\end{equation}
These coordinates have been chosen in order to remove the 
damping term in the equations of motion, and all dependence on the 
expansion now appears only in a time dependent force term, of which 
the amplitude decays as $1/ \tilde{t}^2$. 
At sufficiently long times, therefore, 
an isolated subsystem of the infinite 1D ``expanding" model has, to an 
arbitarily good  approximation, the equations of motion of a strictly 
isolated non-expanding 1D self-gravitating system.  We 
will refer to the coordinates in (\ref{physical coordinates}) as
{\it quasi-physical coordinates}: they are the closest analogy
we have to what are called physical coordinates in the 
three dimensional problem (i.e. $\vec{r} = a(t) \vec{x}$, $t$).
In these coordinates the 3D cosmological equations of motion
are equivalent to those for an infinite Newtonian self-gravitating 
system expanding about a centre in an infinite (static) space.
Correspondingly, any subsystem (e.g. galaxy) taken 
far from other mass in an expanding universe will
evolve in these coordinates like an isolated Newtonian
system (see e.g. \cite{joyce+syloslabini_2012} for a derivation).
The additional term in  $1/ \tilde{t}^2$ in the 1D model,
which we cannot get rid of by a coordinate transformation,
appears because we have imposed in the derivation of
the model the 3D expansion law rather than using
the analogy of this law in 1D (which gives a
collapse in a finite time, qualitatively different to
that of the 3D expansion which we wish to
model).
It is important to note that, re-expressed in terms of 
the three dimensional scale factor $a(t)$, 
the transformation in (\ref{physical coordinates}) 
from the ``comoving'' coordinates $y_i$ to 
the ``quasi-physical'' coordinates in 
(\ref{physical coordinates}) is $r_i= e^{2 \Gamma \tau/3}y_i \equiv a^{1/3} y_i$,
and {\it not} $r_i= a y_i$ as one might naively
expect\footnote{We note that  \cite{yano+gouda} give an incorrect 
generalization of the stable clustering hypothesis 
to the 1D EdS model assuming that stability will be attained 
in the coordinates $a x$. As a result they arrive at
the incorrect conclusion from their numerical study 
that stable clustering is not observed in the model.}.

\subsection{Scaling of a finite virialized subsystem}

The second term on the right hand 
side of Eq.~(\ref{1Deom-finite-finale}) may 
be rewritten simply as $[\frac{\kappa^2}{9}+ 1] g n_0 y_i$
The first term is of order 
$gN \sim g n_{\cal{S}} L$ where $n_{\cal{S}}$ is 
the average comoving density of the subsystem.
Thus if we consider any isolated 
subsystem which is already highly 
non-linear (with $n_{\cal{S}} \gg n_0$) 
we are necessarily in the regime where
the second term may be neglected.
In this case the isolated subsystem will
thus behave, in quasi-physical coordinates,
like a truly isolated 1D self-gravitating system.
Just as in 3D, such a system, for large $N$, 
in the collisionless limit, attains on a 
few dynamical time scales ($\sim 1/\sqrt{gn_{\cal{S}}}$)
a stationary virialized state (see e.g. 
\cite{joyce+worrakitpoonpon_2012} and 
references therein), i.e., a system which 
becomes time independent in the 
quasi-physical coordinates $r_i$.
More generally, even if not strictly virialized,
it rapidly attains a well defined average size
about which virial oscillations persist.

\subsection{Linear amplification and self-similarity}
In order to derive the exponent in what follows we will
also use the self-similar behaviour of the evolution, which,
for the two point correlation function $\xi(x,\tau)$ means 
that 
\begin{equation}
\xi(x, \tau)=\xi_0 \Big(\frac{x}{R_s(\tau)} \Big)
\end{equation}
in the spatial range where the self-similarity is observed
\footnote{We follow the standard notation here in which
$\xi (x)$ denotes the reduced two point correlation function,
i.e.,  $\xi (x) = n_0^{-2} \langle n(0) n(x) \rangle -1$
where $n(x)$ is the local particle density and $\langle \cdots \rangle$
denotes the average over the statistically
translationally invariant point process. See e.g. \cite{peebles}
or \cite{book}.}.
The function $R_s(\tau)$ is derived from linear theory,
which describes a self-similar evolution from power
law initial conditions once the growing mode dominates.
In the 1D model the linear theory
growth of perturbations can be easily derived by
using the expression for the force Eq.~(\ref{1D-exact-force-disp})
in terms of the displacements $u_i$. 
Setting, without loss of generality, the mean 
displacement  $\langle u \rangle$ to zero,
we have simply 
\begin{equation}
\frac{d^2 {u}_i}{d\tau^2} +
\Gamma \frac{d { u}_i}{d\tau} = 2g n_0 u_i \,.
\label{linear theory}
\end{equation}
which is characterized by the growing mode 
\begin{equation}
{u}_i \propto e^{2 \alpha \Gamma \tau}=a^{\alpha}\,  \quad {\rm where} \quad
 \alpha = \frac{1}{4}[-1 + \sqrt{1+\frac{24}{\kappa^2}}]\,.
 \label{linear theory-growing}
\end{equation}
For $\kappa=1$ we thus recover the same growth law as in 
the standard 3D EdS model.
Assuming a power law power spectrum of density fluctuations
$P(k) \propto k^n$ (and thus $\xi \propto 1/x^{n+1}$ at large $x$)
the evolution is self-similar with
\begin{equation}
R_s = e^{\frac{4 \alpha}{1+n} \Gamma \tau}=a^{\frac{2 \alpha}{1+n}}\,.
\label{SS-scaling}
\end{equation}

\subsection{Exponent of non-linear two point correlation function}

\begin{figure}
\vspace{1cm}
{
\par\centering \resizebox*{9cm}{8cm}{\includegraphics*{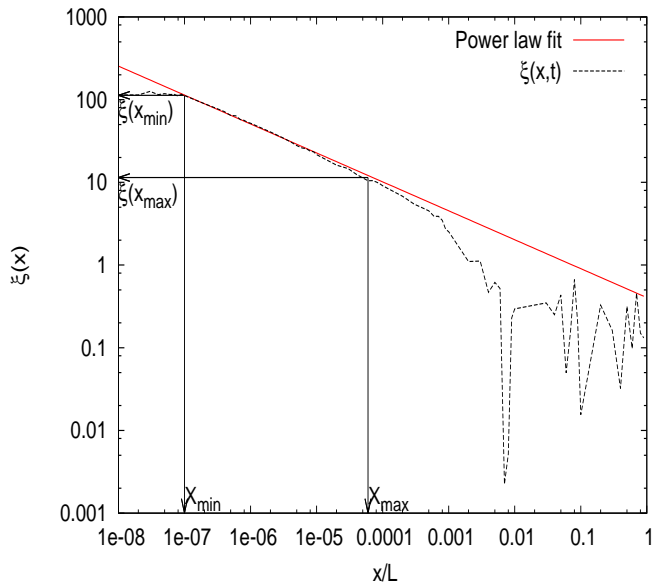}}
\par\centering
}
\caption{Two point correlation function measured in an evolved simulation, with $\kappa=1$ and $n=2$.
The values of $x_{min}$ and $ x_{max}$ define the range in which there is (by eye) an apparent power-law 
behaviour.} 
\label{fig1} 
\end{figure}

Shown in Fig.~\ref{fig1} is a typical two point correlation $\xi(x)$
obtained from the evolution of power law initial conditions in the class
of models we study. Between a lower cut-off scale  $x_{min}$, and
an upper cut-off scale $x_{max}$ a simple power law behaviour of
the correlation function is observed $\xi \sim x^{-\gamma}$.
The exponent $\gamma$ may be written as 
\begin{equation}
	\gamma  =  - \frac{{\ln \left( {{\xi \left(x_{\max}\right)}} \right) - \ln \left( {{\xi \left(x_{\min}\right)}} \right)}}{{\ln \left( {{x_{\max }}} \right) - \ln \left( {{x_{min}}} \right)}}\,.	
\end{equation}

The temporal evolution of the correlation function is found to
be self-similar, down to the scale $x_{\min}$, below which
it is broken. The scale $x_{\max}$ is in the range in
which self-similarity applies, and thus scales in proportion
to $R_s$, with $\xi \left(x_{\max}\right)$ a fixed amplitude corresponding
to the transition between the linear and non-linear regime. 

If we make now the hypothesis that the structures 
giving rise to the correlation measured at the lower
cut-off are {\it stable in quasi-physical coordinates},
i.e., these structures behave like isolated 
approximately virialized structures, 
we have that $x_{min} \sim a^{-1/3}$.
Further, since for $\xi \gg 1$, $\xi(x)$ 
mesures simply the average density at 
distance $x$ from an occupied point 
normalized to the mean density, we
have that $\xi \left(x_{\min}\right) \sim a^{1/3}$. 
It follows that the exponent tends 
asymptotically (at sufficiently long 
times) to 
\begin{equation}
	\gamma_{\rm sc} \left( {n, \kappa } \right) = \frac{{2\kappa \left( {n + 1} \right)}}{{\kappa \left( {2n - 1} \right) + 3\sqrt {{\kappa ^2} + 24} }}
\label{sc-prediction}
\end{equation}
This result is identical to that derived in \cite{joyce+sicard_2011}
using more heuristic arguments based on the scaling of
the energy\footnote{In \cite{joyce+sicard_2011}, as in 
various other papers on these models, units were used 
in which $2gn_0=1$, and thus $\Gamma=\kappa/ {\sqrt 6}$.}. It tells us how the exponent,
in the stable clustering hypothesis, depends on the 
initial conditions, characterized by the exponent $n$
of the power spectrum of density fluctuations, and
on the expansion rate parametrized by the constant
$\kappa$, which we recall is 
\begin{equation}
\kappa= \Gamma \sqrt{3/gn_0}= \left(\frac{3H^2 a^3}{8\pi G \rho_0}\right)^{1/2} \,.
\end{equation} 
Note that as $\kappa \rightarrow 0$ we obtain 
$\gamma_{\rm sc}=0$, i.e. a correlation function which
is predicted to be flat over an arbitrarily large
distance. Indeed, in this static limit, the
stable clustering hypothesis corresponds to
stability also in comoving coordinates, and
therefore to a constant $x_{min}$ 
and $\xi \left(x_{\min}\right)$.  Clearly, however, we
do not expect in this case that the formation of 
non-linear structure at a given scale will leave 
approximately unperturbed the structures
formed previously at smaller scales. 

\section{Numerical simulations: methods and results}

\subsection{Method and calibration}

The numerical study in \cite{joyce+sicard_2011}, as 
most previous studies of the three models belonging to 
the class we are considering, exploits fully the very 
attractive property of these models that they 
admit ``exact" numerical integration:
the force on particles, whether given in the form 
(\ref{1D-exact-force-disp}) or (\ref{1D-force-periodic}),
is such that the motion (with or without damping) admits 
an analytical solution other than at particle crossings.
An event driven algorithm can then be employed,
in which the determination of the time of the 
next ``collision", and which 
pair of particles it involves, requires only the 
solution of algebraic equations. As described 
by \cite{noullez_etal} the integration can be 
sped up optimally using a ``heap" structure. 

For the previously considered cases, corresponding
to $\kappa=0,1,\sqrt{3}$, the algebraic equations
involved are, respectively, quadratic, quintic and
cubic. For a generic $\kappa$, however, the 
equations are not polynomial, and the numerical 
cost of solving them increases and makes the code,
which is already expensive numerically, even more
time consuming\footnote{The equation for the
crossing times is of the form $1+Az^\beta+
Bz^{(\beta+1)/2}=0$ where $z=e^{\Gamma \tau}$,
$\beta=\sqrt{1+\frac{24}{\kappa^2}}$, and $A$ and
$B$ are constants.}. As we wish to study 
fully a broad range of $\kappa$ and initial conditions,
we have chosen here instead to use a particle-mesh  
(PM) code of which the numerical speed can be greatly 
enhanced by the use of FFT techniques and 
parallelization\footnote{See, e.g., http://astro.uchicago.edu/~andrey/Talks/PM/pm.pdf
and references therein.}. In order to be sure that we are 
not, as a result, losing the advantages of the precision 
of the integration accessible in these 1D models, 
we calibrate, as discussed below, our choice of the 
new discreteness parameters introduced by the PM code 
by comparing our results with those obtained, 
for $\kappa=0,1,\sqrt{3}$, using the ``exact" code.  

Our PM code defines a regular grid on the simulation box,
on which a density field $n(x)$ is defined at each time step
by a ``clouds in cell" interpolation of the particle
positions. The modified 1D Poisson equation
\begin{equation}
\frac{d ^2 \phi}{d^2 x}= 2g [n(x)-n_0]
\end{equation}
is then solved for the potential in Fourier space using an FFT, using the solver 
fftw3 \footnote{See http://www.fftw.org/}. An inverse FFT then determines 
the force on the grid, which is then interpolated back onto the particle positions.
The latter are then advanced using a leapfrog algorithm, 
with a simple adaptative timestep: it is chosen at
each time by imposing that the mean distance travelled by  
particles during a time step be equal to a chosen fraction 
$\eta$ of the PM grid. 

\begin{figure}
{
\par\centering \resizebox*{9cm}{8cm}{\includegraphics*{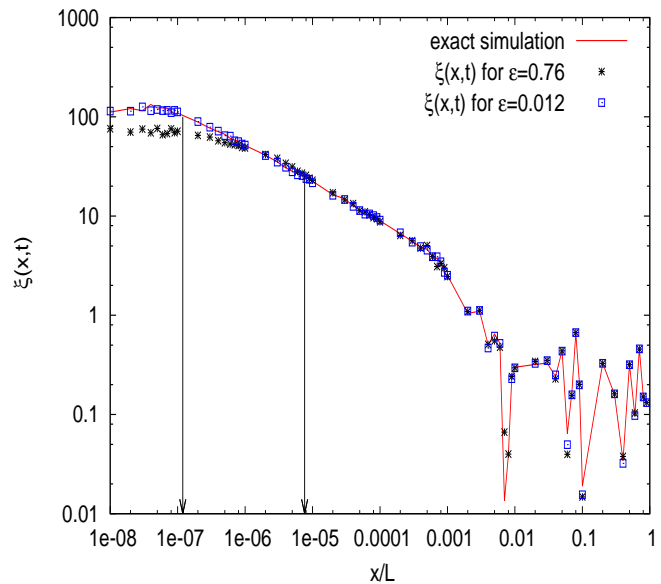}}
\par\centering
}
\caption{Two point correlation functions obtained by evolving from 
the same initial condition (with $n=2$) and $\kappa=1$, for three 
different cases: using the ``exact" code (solid line) and using our 
PM code with the two indicated values of the PM grid spacing.}
\label{fig2}
\end{figure}

Compared to the exact code, our PM code therefore introduces
resolution effects controlled by two dimensionless parameters:  the
size of the PM grid compared to the lattice spacing
in the initial conditions, which we denote by $\epsilon$, 
and the parameter $\eta$ controlling the size of 
the time steps. We have made 
an extensive study of the effects introduced by this finite 
resolution of the PM code, and in particular  on the determination 
of the two point correlation function which is the quantity which 
interests us here. Shown in Fig.~\ref{fig2} is, for example, a 
comparison of the results for runs from identical initial conditions, 
for the case  $n=2$ and $\kappa=1$, obtained from the exact
code and the PM code, for two different values of $\epsilon$,
and $\eta$ sufficiently small (of order unity) so that  convergence
is obtained with respect to it. We find, very reasonably, that excellent 
agreement is obtained provided $\epsilon$ is chosen smaller than 
$x_{min}$, while a larger $\epsilon$ than this minimal scale leads
to a visible deviation of the correlation function
from its correct value at scales below $\epsilon$.
We have chosen our numerical parameters in the simulations 
reported below in order to obtain such an indiscernible difference in the 
correlation function between our code and the exact code, for the cases 
$\kappa=0,1,\sqrt{3}$. All the results presented here are for systems 
with $N=10^5$ particles (in periodic boundary conditions).

For initial conditions with power law power spectra
$P(k) \sim k^n$ (at small $k$, a cut-off at large
$k$ is always implicitly assumed), we expect 
hierarchical structure formation in $d$ dimensions 
for $n$ in the range $-d<n \leq 4$ \citep{peebles}: 
for $n \leq -d$ the variance of mass fluctuations
diverges at large scales, while for $n > 4$ 
linear theory breaks down (and the dynamics
at large scales will be determined by the
power at the ultra-violet scale). We thus
consider (as in \cite{joyce+sicard_2011})
the cases $n=0, 2, 4$. Point processes
with such power spectra are easy to
set up: $n=0$ is obtained by randomly
distributing points in the interval (giving
a power spectrum $P(k) = 1/n_0$), the
case  $n=2$ by applying small random 
and uncorrelated displacements to a regular 
lattice, and $n=4$ in the same way 
but with the additional constraint that 
neighbouring pairs have equal and 
opposite displacements 
(see \cite{gabrielli+joyce_2008} and
references therein). 

\subsection{Results}

\begin{figure}
{
\par\centering \resizebox*{9cm}{8cm}{\includegraphics*{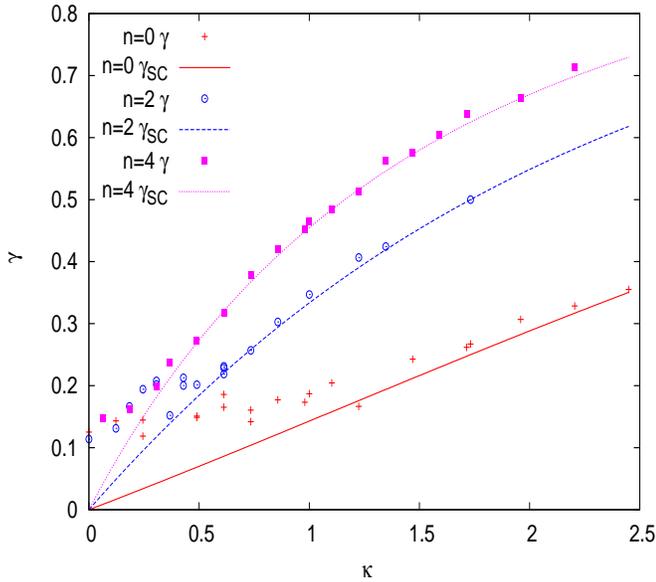}}
\par\centering
}
\caption{Exponents of non-linear power law clustering $\gamma$ 
measured in simulations with different initial conditions 
(parametrized by $n$) and expansion rates (parametrized by $\kappa$).
The solid lines are the predictions of stable clustering, Eq.~(\ref{sc-prediction}).}
\label{fig3}
\end{figure}

Our results are summarized in Fig.~\ref{fig3}. Each point on the plot
corresponds to a single simulation with given $n$ and $\kappa$, where
the latter varies in the range $0$ ---$ 2.5$. The figure shows then the
value of the exponent $\gamma$ of the power law
region of the correlation function. In each case the given value
is determined using a linear regression in the range 
between $x_{min}$ and $ x_{max}$, the latter being determined 
themselves by eye (cf. Fig.~\ref{fig1}). 
The uncertainty in the fitted exponent associated with 
the (by eye) estimation of the end-points for the fit 
is very small (of order a few percent at most)
in most parts of the parameter space. However, as
we will now discuss and explain, 
at lower values of $\kappa$ and $n$, where the extent
of the power law region is much smaller, the resultant
uncertainty becomes more significant. The continuous curves 
plotted correspond to the predictions of the stable clustering 
hypothesis, Eq.~(\ref{sc-prediction}).  

\begin{figure}
{
\par\centering \resizebox*{9cm}{8cm}{\includegraphics*{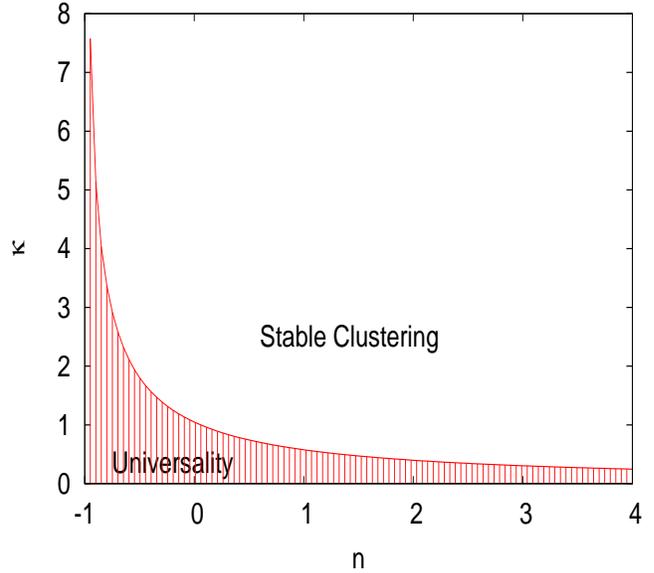}}
\par\centering
}
 \caption{Schematic representation of our results: the space
 of initial conditions ($n$) and cosmology ($\kappa$) breaks
 into two parts, with a boundary defined approximately
 by $\gamma_{sc} (n, \kappa)= 0.15$.}
\label{fig4}
\end{figure}

Our results show excellent agreement with the stable
clustering prediction for a very large part of the 
explored parameter space of initial conditions 
and cosmology. This region can be well 
characterized simply by the condition 
$\gamma_{sc} (n, \kappa) \gtapprox 0.2$,
i.e., good agreement with stable clustering
is observed simply in the region where
the predicted exponent be larger than 
some critical value. In the rest
of the parameter space (i.e. for
$\gamma_{sc} (n, \kappa) \ltapprox 0.2$)
there is clear disagreement with the stable
clustering prediction, and the
measured exponents lie in a very 
narrow range, between $0.2$ and 
$0.15$, or a little smaller. It is precisely 
in this part of the parameter space, 
however, that there is also a 
considerable scatter about the 
curves of the measured
exponents. This scatter is in
fact just a measure of the 
uncertainty in the determination
of these exponents, which is
most difficult (for reasons
we explain in detail below) in 
the region where the exponents
become small. Thus, our results
are quite consistent with 
the hypothesis that the exponent in the region 
 $\gamma_{sc} (n, \kappa) \ltapprox 0.2$
is ``universal", in the sense that it is 
independent of initial conditions and 
cosmology. A schematic representation
of this result is shown in Fig. ~\ref{fig4}: the 
parameter space of  initial conditions
and cosmology breaks up into a ``stable clustering" 
region, and a region of ``universality".
The contour defining the two disjoint regions
is simply given by 
$\gamma_{sc} (n, \kappa) =\gamma_0$.
Rather than a line, there may of course
be a small region in parameter
space in which the exponent varies
in a continuous manner from its
stable clustering value to a limiting
universal value of $\gamma_0 \approx 0.15$
\footnote{In the language of statistical physics
our result could be described as an evidence
for  an ``out of equilibrium phase transition": as control 
parameters for the system are changed the 
(out of equilibrium) behaviour of the system
changes in a discontinuous manner. With
the precision of our current results we cannot
determine with confidence whether 
there is really such a transition or rather a
continuous ``crossover'' between the two
``phases".}.

\section{Discussion and conclusions}

\subsection{Interpretation of 1D results}
We have found numerically that the stable clustering 
hypothesis predicts the exponents of non-linear
clustering very accurately at sufficiently large
$n$ and $\kappa$, i.e., if the spectrum of initial
fluctuations is ``sufficiently blue" and the expansion
rate is sufficiently fast. More precisely it appears that
the criterion for it to work is that $\gamma_{\rm sc}$,
which is a monotonically increasing function of both
$n$ and $\kappa$, be larger than some critical value.
It is not difficult to give a simple physical explanation
for this behaviour, as we now explain.

Let us consider two overdense regions,  initially 
(in the linear regime) of comoving size $L_1^0$ and 
$L_2^0 > L_1^0$. Self-similarity of their evolution 
implies that there is simply a time delay 
$\Delta \tau_{12}$ between every stage of their 
evolution given by 
$R_s(\Delta \tau_{12}) =(L_2^0/L_1^0)$.
Using Eq.~(\ref{SS-scaling}) we obtain 
\begin{equation}
\Delta \tau_{12}=\frac{1+n}{4\Gamma \alpha} \ln (L_2^0/L_1^0)\,.
\label{time-delay}
\end{equation}
Let us suppose that the fluctuation $1$
goes non-linear and approximately virializes 
at some given time $\tau_1^v$, when its size is $L_1^v$.
When fluctuation $2$ virializes, at time 
$\tau_2^v=\tau_1^v+\Delta \tau_{12}$, its
size is $L_2^v=(L_2^0 / L_1^0) L_1^v$.
If in the interval between $\tau_1^v$
and $\tau_2^v$, structure $1$ behaves 
to a very good approximation as an 
isolated structure, its size in comoving
coordinates will decrease by a factor
$e^{-2\Gamma \Delta \tau /3}$.
It follows that the relative size of
the two structures at the time when
$2$ virializes is
\begin{equation}
\left(\frac{L_2}{L_1}\right)=\left(\frac{L_2^0}{L_1^0}\right)^{-\frac{1+n}{6\alpha}}  \left(\frac{L_2^0}{L_1^0}\right)
\label{scaling}
\end{equation}
where $\alpha$ is the growing mode exponent given in
(\ref{linear theory-growing}).  Quite simply, the faster
the expansion rate, or the larger is $n$,  the more 
a given scale which has entered the non-linear regime
can ``shrink" relative to a larger scale in the time
before the latter goes non-linear. 
Further we note, using  (\ref{linear theory-growing}), that
Eq.~(\ref{scaling}) can be written 
\begin{equation}
\left(\frac{L_2}{L_1}\right)=\left(\frac{L_2^0}{L_1^0}\right)^{-\frac{\gamma_{\rm sc}}{1-\gamma_{\rm sc}}}  \left(\frac{L_2^0}{L_1^0}\right)
\end{equation}
i.e., the dependence on $(n, \kappa)$ is in fact completely
determined through the exponent $\gamma_{\rm sc}$ itself.

The predicted stable clustering exponent is therefore
directly related to a physical quantity which we would expect 
naturally to control the validity of the assumption 
made in deriving it:  the more ``concentrated" are the 
pre-existing virialized substructures inside a larger 
structure when it collapses, the better should become 
the approximation that this structure will behave simply 
as a collection of sub-structures which are not
disrupted by the virialization and subsequent evolution
of the larger structure. Indeed, 
in the limit  that $\gamma_{\rm sc}$ 
approaches unity, any structure which collapses 
and virializes will see the substructures 
it contains essentially as point particles.  
As $\gamma_{\rm sc}$ decreases, on the 
other hand, we expect that the interaction 
between structures can lead to their disruption, 
and in particular that ``mergers" 
of substructures become much more probable.
Given that the value of the exponent 
$\gamma_{\rm sc}$ characterizes precisely
the relative ``condensation" of scales, it
is very natural that a critical value of the
exponent should characterize the 
breakdown of stable clustering. 

What our results do not allow us to conclude, as
we have discussed, is what precisely happens
in the regime where stable clustering breaks 
down: our numerical results do not allow us
to clearly distinguish between the possibility of
an abrupt (discontinuous) transition
to a region in which there is a truly
universal exponent, or a smoother transition
to a region in which the exponent depends 
only very weakly on the relevant parameters
($n, \kappa$).  Further we do not (currently)
have a model to explain the exponent (or
narrow range of exponents) observed
in this part of the parameter space. 

Extending the analysis given just above
to determine the scalings of relative sizes
of structures,  it is simple to understand why 
it is precisely  the region of  small $\gamma_{\rm sc}$ 
in which the numerical results for the exponent
of the power law in the correlation function 
are most noisy. The degree of precision in the 
measured exponent is essentially just a function 
of the range of scale over which the power-law 
behaviour extends, i.e., 
it depends on the ratio $x_{min}/x_{max}$
accessible in a numerical simulation.
The scale $x_{max}$ corresponds
approximately at any time to the size
of the largest (approximately) virialized
scale. Let use denote by $x_i$ the 
comoving size of the {\it first} structure 
which virializes in the simulation, at
the time $\tau_i$ at which it virializes.
Self-similarity and stable clustering then
imply the temporal evolutions
\begin{equation}
\frac{x_{min}}{x_i} \sim e^{-2\Gamma(\tau-\tau_0)/3}\,
\qquad
\frac{x_{max}}{x_i} \sim  \frac{R_s( \tau)}{R_s(\tau_0)} 
\sim e^{-\frac{4\alpha}{n+1} \Gamma(\tau-\tau_0)}\,.
\end{equation}
We can infer, using Eq.~(\ref{time-delay}), that,
the range over which power law scaling is
expected for $\tau > \tau_0$ at the 
end of the simulation can be expressed as 
\begin{equation}
\ln \left (\frac {x_{max}}{x_{min}}\right)= \frac{1}{1-\gamma_{\rm sc}}
\ln \left (\frac {L_f^0}{L_i^0}\right)= \frac{1}{1-\gamma_{\rm sc}}
\ln \left (\frac {N_f}{N_i}\right)
\label{range-power-law}
\end{equation}
where $L_i^0$ and $L_f^0$ are, respectively,
the initial comoving size of the first and last
scale to virialize during the evolution, and
$N_f$ ($N_i$) are the number of particles
they contain, respectively. Thus, while for 
$\gamma_{\rm sc}$ closer to unity this
range of scales is large, as $\gamma_{\rm sc}$
decreases it contracts and is simply of order 
the ratio $\frac {L_f^0}{L_i^0}$ for
$\gamma_{\rm sc} \approx \gamma_0$. 

For our simulations, with $N=10^5$ particles,
taking $N_i \sim 10$, or a little larger, 
and $N_f \sim 10^3-10^4$,
we can, for small $\gamma_{\rm sc}$,
access power law clustering over 
approximately two orders of 
magnitude, which leads in practice to the 
observed order of the uncertainly in 
the exponent ($\sim 0.05$).

\subsection{Comparison with 3D simulations: correlation analysis}

Let us compare our results to those in three dimensions.
Numerical simulations have been done by various groups
to test the stable clustering hypothesis, via the study of
the EdS model starting from power-law initial conditions
(see, e.g. \cite{efstathiou_88, padmanabhan_etal_1996, colombi_etal_1996, jain+bertschinger_1997,
jain+bertschinger_1998, ma+fry_2000c, smith}). In this case, corresponding 
to $\kappa=1$ in our notation, the predicted exponent, in three 
dimensions, of the two point correlation function is  
$\gamma_3=3(3+n)/(5+n)$. The range of power spectra 
which has been probed numerically is $-3<n\leq 1$. 
Further, more generally, the stable clustering hypothesis
leads to scaling relations for higher order correlation
functions, which have been tested by some authors
 \citep{colombi_etal_1996, ma+fry_2000c}.
 
Until the most recent and extensive study of \cite{ smith},
previous works had concluded that the stable clustering
prediction worked quite well, with, in some cases, 
indications of small deviations close to the limits  
of numerical resolution. This last study, on the other
hand, reports measurements of the 
exponent $\gamma$ which are clearly discrepant 
with the stable clustering prediction: for $n=-2$, $\gamma=0.77$
rather than $\gamma_3=1$; for $n=-1.5$, 
$\gamma=0.91$ rather than $\gamma_3=1.29$;
for $n=-1$, 
$\gamma=1.26$ rather than $\gamma_3=1.5$;
 for $n=0$, 
$\gamma=1.49$ rather than $\gamma_3=1.8$.

These results are clearly {\it qualitatively} different 
from those we have found in our 1D models, for
any $\kappa$. While the measured $\gamma$ 
increases as $n$ increases, just as in one
dimension, the discrepancy with stable clustering 
in three dimensions  is associated with an exponent $\gamma$ 
which is {\it smaller} than the  predicted 
stable clustering exponent. Further there
is no indication --- in this limited range of $n$ 
probed --- that the discrepancy with respect
to stable clustering is increasing, nor that
the associated discrepancy manifests itself
as an (exact or approximate) universality
of the measured exponent. 

These differences between 1D and 3D results 
could, of course, be simply a reflection of the
difference of the gravitational clustering in 
the two cases. One difference which may be 
essential, as discussed in \cite{joyce+sicard_2011},
is that in one dimension there are no tidal 
forces, and a virialized structure may only be
perturbed by another structure actually 
crossing it (which is, on the other hand,
much more likely than in three 
dimensions). However, one of the main 
motivations for this study is precisely 
that the reliability of results from 3D 
simulations is not clear: 
the combination of the fact that the 
range of comoving scales probed
grows in proportion to $N^{1/3}$,
and the necessity to introduce a 
(relatively large) smoothing parameter
to regularize the small scale divergence 
in the 3D gravitational force, mean 
that the exponents (in existing studies)
are measured over at most a little more 
than one order of magnitude. 
Further
there are consideable discrepancies between
some of the studies. \cite{jain+bertschinger_1998}
find, for example, good agreement with stable
clustering for the case $n=-2$ with
simulations of a similar size to those
of \cite{smith}\footnote{This notable difference is 
attributed by  the latter, on the basis of a 
qualitative analysis, to the fact 
that \cite{jain+bertschinger_1998} evolve to a 
time at which the amplitude of the power at
the scale of the box is higher, leading possibly
to poor modelling of the coupling to 
the ``missing" long wavelength modes. Another
notable difference in the simulations is, however,
that a considerably smaller smoothing length is
employed by  \cite{jain+bertschinger_1998}.
As discussed below, a larger smoothing
could be responsible for suppressing power on
smaller scales, leading potentially to an 
effective exponent which is lowered.}.

Let us focus a little more on these differences 
between 1D and 3D simulations linked to the small
scale smoothing of the force. In 1D, as discussed, the equations 
of motion can be integrated exactly 
in absence of smoothing, so that there
is in practice {\it no lower limit} on spatial
resolution other than that arising
from the finite particle density, of order
the scale $x_{min}$ in the correlation
function.  We note that, following the 
analysis above, we infer that, if
stable clustering holds, we have
\begin{equation}
\ln \left (\frac {x_{min}}{x_{i}}\right)= \frac{\gamma_{\rm sc}}{1-\gamma_{\rm sc}}
\ln \left (\frac {L_f^0}{L_i^0}\right)
\end{equation}
in one dimension, while the same arguments applied to
three dimensions give
\begin{equation}
\ln \left (\frac {x_{min}}{x_{i}}\right)= \frac{\gamma_{\rm sc}}{3-\gamma_{\rm sc}}
\ln \left (\frac {L_f^0}{L_i^0}\right)
\end{equation}
The scale $x_{i}$, the comoving size of the first structure which virializes,
is of order the initial interparticle spacing $\lambda$. In one dimension,
even when we use, as here,  a PM code, the smoothing can, without 
excessive numerical cost, be chosen so that it is at all times considerably
 smaller than the scale $x_{min}$, so that we can be absolutely confident that 
it plays no role in the evolution. In three dimensions, on the contrary,
cosmological simulations of a reasonable size (i.e., so that the
range of scales $\frac {L_f^0}{L_i^0}$ which go non-linear
cover at least a decade or two) are typically performed with a 
smoothing of between one tenth and one hundredth of the 
interparticle spacing $\lambda$.  Thus 3D simulations cannot
in practice simulate systems manifesting stable clustering only
over a very limited range of length scales, while remaining in the
regime where $\varepsilon \ll x_{min}$. While the use
of a cut-off $\varepsilon \geq  x_{min}$ does not 
necessarily imply that the clustering above the scale
$\varepsilon$ is not accurately reproduced, it is quite
possible that this could be the case if great care is not 
taken in the choice of numerical parameters
(see \cite{knebe_etal_2000, joyce+syloslabini_2012}).
In this respect even when a relatively large smoothing
is employed, two body relaxation may play a role
in disrupting the desired collisionless evolution
of $N$-body simulations, and such effects 
can potentially lead to a ``pollution" of larger scales
than $\varepsilon$ \citep{knebe_etal_2000}. 
In the 1D system, in contrast, there is in fact no two body collisionality
analagous to that in three dimensions, 
and collisional relaxation (of which the
mechanism is not yet fully understood)
is extremely slow compared to that
induced by 3D two body relaxation(see 
\cite{joyce+worrakitpoonpon_2010}
and references therein). Simple 
estimates show that such relaxation
should play absolutely no role in
our 1D simulations, while the same
is not true in typical 3D simulations.
Our study thus suggests
that particular attention should be paid to this point
in the analysis of 3D simulations. 

We note also that simulations of clustering in a 3D
universe have been performed for the case of
a static universe, for the cases $n=0$ \citep{bottaccio2}
and $n=2$ \citep{sl1, sl2, sl3}. The results are in
this case qualitatively very consistent with those observed 
here and in \cite{joyce+sicard_2011} for the static
($\kappa=0$) limit. Just as in the 1D simulations
self-similarity is observed, and further a correlation 
function which appears to be the independent 
of the initial conditions, with a very shallow exponent 
in the inner part, $\gamma \approx 0.3$. 

\subsection{Comparison with 3D simulations: halos and universality}

Finally let us discuss the question of ``universality".
A striking feature of our 1D results is that the measured 
exponent characterising clustering appears to become 
independent of initial conditions and cosmology  when the stable
clustering approximation breaks down --- more specifically
when the exponent $n$ in the initial power spectrum
is sufficiently small for any given expansion rate
parameter $\kappa$. In three dimensions, following notably 
the work of \cite{navarro1, navarro2}, much
emphasis has been placed on a ``universality" 
of cosmological gravitational clustering which 
manifests itself as an apparent independence of
halo profiles of initial conditions and cosmology.
The range over which such a universality
might apply is not, however, clear: studies
such as that of  \cite{knollmann_etal_2008}
show that, even in the EdS cosmology, the
inner exponent of halos may vary sensibly for
power law initial conditions as a function of $n$.

Does the apparent universality we observe in 1D
correspond to a universality in halo profiles?
As discussed in \cite{joyce+sicard_2011}, one
can identify and extract halos by prescriptions
analogous to those used in three dimensions.
However, the objects thus extracted are simply
so ``clumpy", at any scale significantly
above the scale $x_{min}$, that it is 
not possible to approximate the run of density
by a smooth profile: indeed it can be verified 
directly that the power law exponent (over
several decades) in the correlation function
corresponds to a truly scale invariant  
fractal distribution of the matter in the
corresponding range.  If, however, the
extracted halos are smoothed (by hand) over 
a scale much larger than $x_{min}$, the
exponent characterising the decay of
density about the centre of the halo, should
be approximately the same as that measured in the 
correlation function. This is true because
the latter measures, by definition, the mean
density as a function of distance about any 
point, which in a simple fractal type distribution
is (modulo fluctuations) the same about 
any point. Thus, in the 1D model, the universality
we observe here is indeed expected to
be associated to a universality of inner 
exponents of halos. 

Extrapolated to 3D, our results thus suggest
that the relative smoothness of halo profiles is 
an artefact of insufficient spatial resolution, 
and that the minimal scale of substructure
is limited only by such resolution. In this
case, for power law initial conditions,  the 
observed power law slope of the correlation
function should in fact {\it coincide} with the
inner exponent of the measured halo 
profiles. We note that this is completely
different to what is predicted usually
in three dimensions on the basis of a
smooth halo model (see e.g. 
\cite{ma+fry_2000c, ma+fry_2000b, ma+fry_2000a,
yano+gouda_2000}).  Up to now 3D numerical
studies have not, to our knowledge, addressed this
point directly (by comparing measured slopes
in halos and two point correlation functions, in scale
free cosmologies). We note that the results of 
 \cite{knollmann_etal_2008} on inner
halo profiles appear, however, to be very 
consistent with a behaviour completely
analogous to what we have seen in our
1D models: in their study of power 
law initial conditions in an EdS cosmology,
for $n$ varying in a range between
$-0.5$ and $-2.75$,  
inner slopes for halo profiles 
are estimated using different fitting procedures.
For the simplest such procedure it
is found (see Fig. 2) that the measured
exponent varies from a value in
good agreement with stable clustering 
for the largest value of $n$, but 
appears to deviate, as $n$ 
decreases, towards an exponent
which is  {\it larger} than predicted
by stable clustering,  tending 
rapidly towards an  asymptotic 
exponent a little larger than unity,
and consistent 
with that of \cite{navarro2}.  This suggests that in 
the 3D EdS model, the universality of halos
would translate to an insensitivity to 
initial conditions below an $n$
of order $-2$. Given that the range of effective 
exponents of initial conditions of typical
cosmological (e.g. $\Lambda$CDM) initial 
conditions falls in or close to this range, 
this would appear to be very consistent
with the approximate universality of halo 
profile exponents observed in these kind of
simulations.

\subsection{Future studies}

Our 1D study thus motivates further careful numerical study of
non-linear clustering in the pure matter ($\kappa=1$) EdS model 
in three dimensions, extending recent studies
such as \cite{smith} and \cite{knollmann_etal_2008}.
In the same way as has been done here, the study of
this model can be extended to the full family of EdS
models of which one limit is a static universe.
The goal of such a study would be to see if, in this
space of models, and for power law initial conditions, a 
similar structure is found to that we have
seen in one dimension.  Our 1D study suggests
that particular attention needs to be applied to
controlling for the robustness of measure of exponents 
in the two point correlation function to small scale force 
smoothing, particularly when an exponent below the 
predicted stable clustering value is found (as in  \cite{smith}).
Further, study of the relation between exponents measured 
from the two point correlation function and those measured 
from halos,  and their variation as a function of $n$ and $\kappa$, 
should clarify whether the qualitative nature of clustering
is indeed like that usually assumed in theoretical
modelling (smooth halo models) or instead like
that in the 1D models (hierarchical fractal clustering).

If the framework suggested by the 1D models --- 
of  a ``universal" region in the space of initial conditions of clustering, 
delimited by a ``critical" value of the predicted stable clustering
exponent  --- is correct, and applies also in three dimensions, the 
theoretical problem remains of understanding both the corresponding 
value of the ``universal exponent", and why this universality applies 
in the specific region where it does. Further study of the 1D model 
may then help to throw further light on this. As the static model
is in the relevant part of the parameter space, it may suffice to
study this particular limit, i.e., the region in which there is universality
is where the expansion of the universe has little effect
on the non-linear clustering.  We note that a very recent study 
in \cite{schulz_etal_2012}  shows that there is an apparent 
universality in the properties of the ``halos" formed from 
the 1D collapse of a single structure with a range of cold initial 
conditions, and argues that this universality may be related
to that observed numerically in three dimensions. 
The 1D halos in \cite{schulz_etal_2012}, however, are 
smooth, very different to the non-linear structures we observe 
here, which are very clumpy even in the static limit of the 1D 
model. As discussed above, however, in this limit the 
range of non-linear clustering explored by our simulations
is still quite modest, and it may be that larger simulations
might establish a link to the study of  \cite{schulz_etal_2012}.

With respect to this last point --- the effect of varying particle number ---  we note that we 
have not discussed here the question of whether the dynamics of these simulations 
in the non-linear regime is representative of the fluid or Vlasov Poisson limit.
In three dimensions this is a crucial question which has  been the subject of discussion 
and some controversy \citep{splinter_1998,  knebe_etal_2000, power_etal_2002,  discreteness3_mjbm, romeo08},
and one might expect that these 1D models, with their large accessible spatial resolution,
could help to throw light on this question. In N body simulations this question can
in principle be directly probed  by comparing simulations in which the particle number 
N is varied while keeping the lower cut-off scale to the initial density fluctuations fixed,
or, alternatively, by increasing particle density in units of  the grid of  our PM code.
Further, in contrast to three dimensions, it should be feasible to address this question 
much more directly by comparison of the results of the integration of the N body system 
with a direct numerical integration of the Vlasov-Poisson equations themselves. This 
is another  interesting future direction for work on these models.

We are indebted to B. Marcos for numerous discussions and suggestions.
We also thank  S. Colombi, B. Miller, J. Morand, J.-L. Rouet, 
F. Sylos Labini, P. Viot and T. Worrakitpoonpon for useful conversations.


\begin{thebibliography}{}

\bibitem[\protect\citeauthoryear{Aurell \& Fanelli}{Aurell \&
  Fanelli}{2002}]{aurell+fanelli_2002a}
Aurell E.,  Fanelli D.,  2002, Astron. Astrophys., 395, 399

\bibitem[\protect\citeauthoryear{Baertschiger, Joyce, Gabrielli \&
  Sylos~Labini}{Baertschiger et~al.}{2007a}]{sl1}
Baertschiger T.,  Joyce M.,  Gabrielli A.,    Sylos~Labini F.,  2007a, Phys.
  Rev., E75, 021113

\bibitem[\protect\citeauthoryear{Baertschiger, Joyce, Gabrielli \&
  Sylos~Labini}{Baertschiger et~al.}{2007b}]{sl2}
Baertschiger T.,  Joyce M.,  Gabrielli A.,    Sylos~Labini F.,  2007b, Phys.
  Rev., E76, 011116

\bibitem[\protect\citeauthoryear{Baertschiger, Joyce, Sylos~Labini \&
  Marcos}{Baertschiger et~al.}{2008}]{sl3}
Baertschiger T.,  Joyce M.,  Sylos~Labini F.,    Marcos B.,  2008, Phys. Rev.,
  E77, 051114

\bibitem[\protect\citeauthoryear{Binney}{Binney}{2004}]{binney_discreteness}
Binney J.,  2004, Mon. Not. R. Astron. Soc., 350, 939

\bibitem[\protect\citeauthoryear{Bottaccio, Pietronero, Amici, Miocchi,
  Capuzzo~Dolcetta \& Montuori}{Bottaccio et~al.}{2002}]{bottaccio2}
Bottaccio M.,  Pietronero L.,  Amici A.,  Miocchi P.,  Capuzzo~Dolcetta R.,
  Montuori M.,  2002, Physica A, 305, 247

\bibitem[\protect\citeauthoryear{{Colombi}, {Bouchet} \& {Hernquist}}{{Colombi}
  et~al.}{1996}]{colombi_etal_1996}
{Colombi} S.,  {Bouchet} F.~R.,    {Hernquist} L.,  1996, Astrophys. J., 465,
  14

\bibitem[\protect\citeauthoryear{{Efstathiou}, {Frenk}, {White} \&
  {Davis}}{{Efstathiou} et~al.}{1988}]{efstathiou_88}
{Efstathiou} G.,  {Frenk} C.~S.,  {White} S.~D.~M.,    {Davis} M.,  1988, Mon.
  Not. R. Astron. Soc., 235, 715

\bibitem[\protect\citeauthoryear{{Ferreira} \& {Joyce}}{{Ferreira} \&
  {Joyce}}{1998}]{ferreira+joyce_1998}
{Ferreira} P.~G.,  {Joyce} M.,  1998, Phys. Rev. D., 58, 023503

\bibitem[\protect\citeauthoryear{Gabrielli \& Joyce}{Gabrielli \&
  Joyce}{2008}]{gabrielli+joyce_2008}
Gabrielli A.,  Joyce M.,  2008, Phys. Rev., E77, 031139

\bibitem[\protect\citeauthoryear{{Gabrielli}, {Joyce} \& {Sicard}}{{Gabrielli}
  et~al.}{2009}]{agmjfs_pre2009}
{Gabrielli} A.,  {Joyce} M.,    {Sicard} F.,  2009, Phys. Rev. E, 80, 041108

\bibitem[\protect\citeauthoryear{Gabrielli, Sylos~Labini, Joyce \&
  Pietronero}{Gabrielli et~al.}{2005}]{book}
Gabrielli A.,  Sylos~Labini F.,  Joyce M.,    Pietronero L.,  2005, Statistical
  Physics for Cosmic Structures.
Springer

\bibitem[\protect\citeauthoryear{{Jain} \& {Bertschinger}}{{Jain} \&
  {Bertschinger}}{1996}]{jain+bertschinger_1997}
{Jain} B.,  {Bertschinger} E.,  1996, Astrophys. J., 456, 43

\bibitem[\protect\citeauthoryear{{Jain} \& {Bertschinger}}{{Jain} \&
  {Bertschinger}}{1998}]{jain+bertschinger_1998}
{Jain} B.,  {Bertschinger} E.,  1998, Astrophys. J., 509, 517

\bibitem[\protect\citeauthoryear{Joyce, Marcos \& Baertschiger}{Joyce
  et~al.}{2008}]{discreteness3_mjbm}
Joyce M.,  Marcos B.,    Baertschiger T.,  2008, Mon. Not. R. Astron. Soc.

\bibitem[\protect\citeauthoryear{{Joyce} \& {Sicard}}{{Joyce} \&
  {Sicard}}{2011}]{joyce+sicard_2011}
{Joyce} M.,  {Sicard} F.,  2011, M. Not. R. Astron. Soc., 413, 1439

\bibitem[\protect\citeauthoryear{Joyce \& Sylos~Labini}{Joyce \&
  Sylos~Labini}{2012}]{joyce+syloslabini_2012}
Joyce M.,  Sylos~Labini F.,  (to appear) 2012, Mon. Not. R. Astron. Soc.

\bibitem[\protect\citeauthoryear{{Joyce} \& {Worrakitpoonpon}}{{Joyce} \&
  {Worrakitpoonpon}}{2010}]{joyce+worrakitpoonpon_2010}
{Joyce} M.,  {Worrakitpoonpon} T.,  2010, Journal of Statistical Mechanics:
  Theory and Experiment, 10, 12

\bibitem[\protect\citeauthoryear{{Joyce} \& {Worrakitpoonpon}}{{Joyce} \&
  {Worrakitpoonpon}}{2011}]{joyce+worrakitpoonpon_2012}
{Joyce} M.,  {Worrakitpoonpon} T.,  2011, Phys. Rev. E, 84, 011139

\bibitem[\protect\citeauthoryear{Knebe, Kravtsov, Gottl\"ober \& Klypin}{Knebe
  et~al.}{2000}]{knebe_etal_2000}
Knebe A.,  Kravtsov A.,  Gottl\"ober S.,    Klypin A.,  2000, Mon. Not. Roy.
  Astron. Soc., 317, 630

\bibitem[\protect\citeauthoryear{{Knollmann}, {Power} \& {Knebe}}{{Knollmann}
  et~al.}{2008}]{knollmann_etal_2008}
{Knollmann} S.~R.,  {Power} C.,    {Knebe} A.,  2008, Mon. Not. R. Astron.
  Soc., 385, 545

\bibitem[\protect\citeauthoryear{{Ma} \& {Fry}}{{Ma} \&
  {Fry}}{2000a}]{ma+fry_2000a}
{Ma} C.-P.,  {Fry} J.~N.,  2000a, Astrophys. J., 543, 503

\bibitem[\protect\citeauthoryear{{Ma} \& {Fry}}{{Ma} \&
  {Fry}}{2000b}]{ma+fry_2000c}
{Ma} C.-P.,  {Fry} J.~N.,  2000b, Astrophy. J. Lett., 531, L87

\bibitem[\protect\citeauthoryear{{Ma} \& {Fry}}{{Ma} \&
  {Fry}}{2000c}]{ma+fry_2000b}
{Ma} C.-P.,  {Fry} J.~N.,  2000c, Astrophys. J. Lett., 538, L107

\bibitem[\protect\citeauthoryear{{Melott}}{{Melott}}{1982}]{melott_prl_1982}
{Melott} A.~L.,  1982, Phys. Rev. Lett., 48, 894

\bibitem[\protect\citeauthoryear{{Melott}}{{Melott}}{1983}]{melott_1d_1983}
{Melott} A.~L.,  1983, Astrophys. J., 264, 59

\bibitem[\protect\citeauthoryear{Miller \& Rouet}{Miller \&
  Rouet}{2002}]{miller+rouet_2002}
Miller B.,  Rouet J.,  2002, Phys. Rev., E65, 056121

\bibitem[\protect\citeauthoryear{Miller \& Rouet}{Miller \&
  Rouet}{2006}]{miller+rouet_2006}
Miller B.,  Rouet J.,  2006, C. R. Phys., 7, 383

\bibitem[\protect\citeauthoryear{{Miller} \& {Rouet}}{{Miller} \&
  {Rouet}}{2010a}]{miller+rouet_2010a}
{Miller} B.~N.,  {Rouet} J.,  2010a, ArXiv e-prints

\bibitem[\protect\citeauthoryear{{Miller} \& {Rouet}}{{Miller} \&
  {Rouet}}{2010b}]{miller+rouet_2010b}
{Miller} B.~N.,  {Rouet} J.,  2010b, ArXiv e-prints

\bibitem[\protect\citeauthoryear{{Miller}, {Rouet} \& {Le Guirriec}}{{Miller}
  et~al.}{2007}]{miller_etal_2007}
{Miller} B.~N.,  {Rouet} J.,    {Le Guirriec} E.,  2007, Phys. Rev. E, 76,
  036705

\bibitem[\protect\citeauthoryear{Navarro, Frenk \& White}{Navarro
  et~al.}{1996}]{navarro1}
Navarro J.~F.,  Frenk C.~S.,    White S. D.~M.,  1996, Astrophys. J., 462, 563

\bibitem[\protect\citeauthoryear{Navarro, Frenk \& White}{Navarro
  et~al.}{1997}]{navarro2}
Navarro J.~F.,  Frenk C.~S.,    White S. D.~M.,  1997, Astrophys. J., 490, 493

\bibitem[\protect\citeauthoryear{Noullez, Aurell \& Fanelli}{Noullez
  et~al.}{2003}]{noullez_etal}
Noullez A.,  Aurell E.,    Fanelli D.,  2003, J. Comp. Phys., 186, 697

\bibitem[\protect\citeauthoryear{{Padmanabhan}, {Cen}, {Ostriker} \&
  {Summers}}{{Padmanabhan} et~al.}{1996}]{padmanabhan_etal_1996}
{Padmanabhan} T.,  {Cen} R.,  {Ostriker} J.~P.,    {Summers} F.~J.,  1996,
  Astrophys. J., 466, 604

\bibitem[\protect\citeauthoryear{Peebles}{Peebles}{1980}]{peebles}
Peebles P. J.~E.,  1980, {The Large-Scale Structure of the Universe}.
Princeton University Press

\bibitem[\protect\citeauthoryear{{Power}, {Navarro}, {Jenkins}, {Frenk},
  {White}, {Springel}, {Stadel} \& {Quinn}}{{Power}
  et~al.}{2003}]{power_etal_2002}
{Power} C.,  {Navarro} J.~F.,  {Jenkins} A.,  {Frenk} C.~S.,  {White} S.~D.~M.,
   {Springel} V.,  {Stadel} J.,    {Quinn} T.,  2003, Mon. Not. R. Astron.
  Soc., 338, 14

\bibitem[\protect\citeauthoryear{{Romeo}, {Agertz}, {Moore} \&
  {Stadel}}{{Romeo} et~al.}{2008}]{romeo08}
{Romeo} A.~B.,  {Agertz} O.,  {Moore} B.,    {Stadel} J.,  2008, Astrophys. J.,
  686, 1

\bibitem[\protect\citeauthoryear{{Schulz}, {Dehnen}, {Jungman} \&
  {Tremaine}}{{Schulz} et~al.}{2012}]{schulz_etal_2012}
{Schulz} A.~E.,  {Dehnen} W.,  {Jungman} G.,    {Tremaine} S.,  2012, ArXiv
  e-prints

\bibitem[\protect\citeauthoryear{Smith, Peacock, Jenkins, White, Frenk, Pearce,
  Thomas, Efstathiou \& Couchman}{Smith et~al.}{2003}]{smith}
Smith R.~E.,  Peacock J.~A.,  Jenkins A.,  White S. D.~M.,  Frenk C.~S.,
  Pearce F.~R.,  Thomas P.~A.,  Efstathiou G.,    Couchman H. M.~P.,  2003,
  Mon. Not. R. Astron. Soc., 341, 1311

\bibitem[\protect\citeauthoryear{Splinter, Melott, Shandarin \& Suto}{Splinter
  et~al.}{1998}]{splinter_1998}
Splinter R.~J.,  Melott A.~L.,  Shandarin S.~F.,    Suto Y.,  1998, Astrophys.
  J., 497, 38

\bibitem[\protect\citeauthoryear{Valageas}{Valageas}{2006}]{valageasOSC_2}
Valageas P.,  2006, Phys. Rev.., E74, 016606

\bibitem[\protect\citeauthoryear{Yano \& Gouda}{Yano \&
  Gouda}{1998}]{yano+gouda}
Yano T.,  Gouda N.,  1998, Astrophy. J. Supp., 118, 267

\bibitem[\protect\citeauthoryear{{Yano} \& {Gouda}}{{Yano} \&
  {Gouda}}{2000}]{yano+gouda_2000}
{Yano} T.,  {Gouda} N.,  2000, Astrophys. J., 539, 493

\end{thebibliography}

\end{document}